\documentclass[twocolumn,twocolappendix,usenames,dvipsnames]{aastex63}
\usepackage{capt-of}

\usepackage{verbatim}
\usepackage{amsmath}
\usepackage{multirow}
\usepackage{relsize}
\usepackage{graphicx}
\usepackage{graphicx}

\setcitestyle{aysep={}} 
\shorttitle{Protostellar Feedback and the N-PDF}
\shortauthors{Kiihne et al.}

\begin{document}

\title{Fitting Probability Distribution Functions in Turbulent Star-Forming Molecular Clouds}

\author{Avery Kiihne}
\email{kiihne1999@gmail.com}
\affiliation{Department of Physics and Astronomy, 
Rutgers University,
136 Frelinghuysen Rd., 
Piscataway, NJ 08854, USA}

\author[0000-0002-6593-3800]{Sabrina M. Appel}
\affiliation{Department of Physics and Astronomy, 
Rutgers University,
136 Frelinghuysen Rd., 
Piscataway, NJ 08854, USA}

\author{Blakesley Burkhart}
\affiliation{Department of Physics and Astronomy, 
Rutgers University,
136 Frelinghuysen Rd., 
Piscataway, NJ 08854, USA}
\affiliation{Center for Computational Astrophysics, 
Flatiron Institute, 
162 Fifth Avenue, 
New York, NY 10010, USA}

\author{Vadim A. Semenov}
\affiliation{Center for Astrophysics, 
Harvard \& Smithsonian, 60 Garden St., Cambridge, MA 02138, USA}

\author{Christoph Federrath}
\affiliation{Research School of Astronomy and Astrophysics, 
The Australian National University, 
Canberra, ACT 2611, 
Australia}

\begin{abstract}
    We use a suite of 3D simulations of star-forming molecular clouds, with and without stellar feedback and magnetic fields, to investigate the effectiveness of different fitting methods for volume and column density probability distribution functions (PDFs). 
    The first method fits a piecewise lognormal and power-law (PL) function to recover PDF parameters such as the PL slope and transition density. The second method fits a polynomial spline function and examines the first and second derivatives of the spline to determine the PL slope and the functional transition density.  We demonstrate that fitting a spline allows us to directly determine if the data has multiple PL slopes. The first PL (set by the  transition between lognormal and PL function) can also be visualized in the derivatives directly. 
    In general, the two methods produce fits that agree reasonably well for volume density but vary for column density, likely due to the increased statistical noise in column density maps as compared to volume density. We test a well-known conversion for estimating volume density PL slopes from  column density slopes and find that the spline method produces a better match ($\chi^2$ of 2.38 vs $\chi^2$ of 5.92), albeit with a significant scatter. Ultimately, we recommend the use of both fitting methods on column density data to mitigate the effects of noise. 

\end{abstract}


\section{Introduction}  \label{sec:intro}

Star formation primarily occurs in dense and cold molecular gas within galaxies. Such gas resides in molecular cloud (MC) complexes and is subjected to the competing effects of supersonic turbulence, self-gravity, magnetic pressure and tension, and feedback from newly formed stars, as well as energetic processes that alter the gas temperature and chemistry \citep[e.g.,][]{Padoan1997,KrumholzBurkhart2018}. 
Due to the complexity of star-forming environments, a common approach for analytic models of star formation is to study the distribution of gas via a density probability distribution function ($\rho$-PDF) analysis \citep[e.g.,][]{FederrathKlessen2012,Burkhart2018}. 
The $\rho$-PDF (volume density) has been used to predict a wide range of star formation observables such as the core mass distribution, the stellar initial mass function \citep[e.g.,][]{Padoan02a, HennebellChabrier2011,Hopkins2012}, the star formation rate \citep[SFR; e.g.,][]{KrumholzMcKee2005, PadoanNordlund2011, FederrathKlessen2012,HennebellChabrier2011}, and star formation efficiency \citep[SFE; e.g.,][]{Federrath+Klessen2013}. 
In addition, models of variable SFE motivated by these works have been used as subgrid star formation prescriptions in galaxy formation simulations to account for the (unresolved) dense gas fraction and its dynamical state \citep{Braun2015,Semenov2016,Li2017,Trebitsch2017,Lupi2018,Gensior2020,Kretschmer2020,Kretschmer2021,Olsen2021}. 
The $\rho$-PDF of atomic and molecular tracers has also been suggested as a diagnostic for the HI-H$_{2}$ transition in a turbulent medium \citep[e.g.,][]{Burkhart2015,Imara2016}

The shape of the gas $\rho$-PDF is profoundly linked to the kinematics and star formation activity of a MC \citep{Burkhart2018, BurkhartMocz2019}. 
The shape of the $\rho$-PDF in MCs is expected to be lognormal (LN) when isothermal supersonic turbulence dominates the gas dynamics \citep{Federrath2008, Passot1998}. Furthermore, the width of the lognormal distribution in the PDF can be related to the sonic Mach number of the gas and the driving mode of the turbulence (compressive or solenoidal) in an isothermal cloud \citep[i.e., the sonic Mach number-variance relation; see e.g.,][]{Federrath2008,Molina2012,Burkhart2018}. 
Generally, the gas within the lognormal portion of the density PDF is controlled by turbulent and magnetic support and is not actively collapsing \citep{Chen2017}. 
The lognormal form of the $\Sigma$-PDF (column density) describes the behavior of diffuse HI and ionized gas as well as some molecular clouds that are not forming massive stars \citep{Hill2008,burkhart10,Kainulainen13a,Schneider+2015,Burkhart2015,Kritsuk+2011,Collins12a,Burkhart2018,Khullar2021} and tends to correspond to extinctions greater than $A_v >1$ or around $n>10^3\;$cm$^{-3}$ \citep{MyersP2015,schneider15,MyersP2017,Alves2017AA,Kainulainen2017,Chen2018}. 

The lognormal plus power-law (LN+PL) form of the gas PDF in MCs has been seen in both observations via column density PDFs ($\Sigma$-PDFs hereinafter) and in numerical simulations \citep[both in density and projected column density; see e.g.,][]{VazquezSemadeni2001, Wada07a,Ossenkopf-Okada2016, Veltchev2019}. 
Several observational studies have shown that there is a strong relationship between the power-law slope of the PDF and the number of young stellar objects \citep[YSOs;][]{Stutz2015A&A...577L...6S,Gutermuth11a,Schneider+2015,Chen2017}.
The power-law slope shallows significantly in less than the mean free fall time \citep{Collins12a,burkhartcollinslaz2015,Federrath2015MNRAS.448.3297F,Guszejnov2018}, while the lifetimes of GMCs are typically between $2$--$10$ free fall times due to disruption \citep{Palla00a,Meidt2015,jefferson2018,Chevance2020,Kim2022,Jeffreson2023}. 
Therefore, the power-law slope (for both the $\Sigma$- and $\rho$-PDFs) possibly traces the accelerated SFR during the early phases of star formation \citep{Murray11b,Lee2015ApJ...800...49L,Lee2016,Grudic+2018}. 
The density at which the PDF transitions to a power-law form heralds the onset of collapse as the self-gravity of the gas dominates at higher densities in the cloud.
Furthermore, stellar feedback promotes gas cycling between the two states of gas (star-forming and non-star-forming) within star-forming clouds and allows the overall star formation efficiency to remain low  \citep{Semenov2017,Appel2022,Rosen_2019}.

A challenge for testing the predictions of star formation models based on the $\rho$-PDF is that the $\rho$-PDF is extremely difficult to observe and its properties usually must be inferred using the $\Sigma$-PDFs \citep{Federrath+Klessen2013, JaupartChabrier2020}. 
For example, one approach includes decomposing the 2D column density data into a set of hierarchical 3D structures: the column density maps of the clouds are decomposed using a wavelet filtering to quantify structure at different spatial scales, first proposed by \citet{Brunt10c} and later refined in \citet{Kainulainen2014}.  

However promising these reconstruction methods are, there are still uncertainties in determining the exact shape of the diffuse lognormal $\Sigma$-PDF (e.g., foreground and background effects, noise), although the power-law slopes and transitional column density are likely robust \citep{Lombardi2015AA, schneider15, Ossenkopf-Okada2016}. 
In addition, model fitting for the shape of the PDF can be affected by biases in the type of fitting used, i.e., assuming the shape of the distribution to be lognormal only (LN), lognormal with a power-law (LN+PL), or lognormal with a power-law and then a second power-law (LN+PLPL). 
The results of fitting the shape of the PDF can also be influenced by the choice of binning when constructing the PDF. 

In this paper we explore using a polynomial spline to fit the PDF, in addition to a more commonly used fitting method, to measure the power-law slope and the transition densities of simulated $\Sigma$-PDFs. 
The advantage of a polynomial spline fit is that the derivatives of the spline fit provide a model-independent mechanism for determining which portions of the density PDF are consistent with a lognormal distribution vs.~a power-law distribution. 

This paper is organized as follows: in Section~\ref{sec:sim}, we describe the numerical simulations used for our analysis; in Section~\ref{sec:models}, we describe each of our fitting methods; and in Section~\ref{sec:results}, we present our $\rho$-PDFs and $\Sigma$-PDFs with each LN+PL fit.
Finally, we discuss our results in Section~\ref{sec:discussion}, followed by our conclusions in Section~\ref{sec:conclusions}.

\section{Simulations} 
\label{sec:sim}

We use a suite of three hydrodynamical simulations performed using FLASH, a publicly available adaptive mesh hydrodynamics code \citep{Fryxell2000}. 
All three simulations are described in \cite{Appel2022}, and the first two were introduced in \cite{Federrath2015}.

FLASH solves the fully compressible MHD equations using adaptive mesh refinement (AMR) and can include many inter-operable modules.
Our simulations use a second-order accurate, Godunov-type method with a 5-wave approximate HLL5R Riemann solver \citep{Waagan+2011}.

Each of these three simulations sequentially includes additional physical processes relevant to star-forming molecular clouds.
The first simulation (denoted as \texttt{Turbulence}) includes driven turbulence and self-gravity but does not include magnetic fields or outflow feedback. 
The second simulation (\texttt{B-Fields}) includes driven turbulence, self-gravity, and  magnetic fields. 
The final simulation (\texttt{All + Outflow}) is identical to the \texttt{B-Fields} run but adds stellar feedback in the form of protostellar outflows.

All three simulations are initialized with uniform density of $\rho_{0} = 3.28 \times 10^{-21}$ g cm$^{-3}$ and a box size of 2 pc, corresponding to a total initial gas mass of $M=388 \, $M$_{\odot}$.  
Turbulence is driven for two turnover times ($t_{\rm turnover} \approx 0.98$ Myr) to fully establish the turbulent cascade before self-gravity and protostellar outflows (as relevant) are initialized. 
Turbulence is then driven continuously throughout the time of the simulation.
Turbulence is driven at half the box size, while at smaller scales it is allowed to develop self-consistently. 
The turbulence driving is set such that the simulations have a velocity dispersion of $\sigma_{v} \approx 1$ km s$^{-1}$ and a sonic Mach number of $\mathcal{M}_s \approx 5$. 
A natural mixture of forcing modes is used, corresponding to an effective driving parameter of $b\approx0.4$ \citep[for more detail, see the driving method in][]{Federrath2010,federrath2022}.
The two simulations with magnetic fields have an initial uniform field strength of $B = 10 \, \mu$G, an Alfv\'{e}n Mach number of $\mathcal{M}_s = 2.0$, and a plasma beta parameter (representing the ratio of thermal and magnetic pressure) of $\beta = 0.33$ \citep{Federrath2015}.
The initially uniform field then distorts and tangles as turbulence develops \citep{lazarianvishniac99,Steinwandel2022}.

Each of the simulations uses sink particles to account for the formation of stars, as described in \cite{Federrath10a, Federrath+2014} and \cite{Federrath2015}. 
Sink particles form only in regions of maximum refinement, where the local gas has undergone gravitational collapse and exceeded the threshold density, $\rho$\textsubscript{sink}.
This threshold is $s$\textsubscript{sink}$=\ln(\rho$\textsubscript{sink}$/\rho_0)=8.75$ for the \texttt{Turbulence} and \texttt{B-Fields} simulations. The \texttt{All + Outflow} case has a higher maximum resolution and therefore a higher sink formation threshold of $s$\textsubscript{sink}$=\ln(\rho$\textsubscript{sink}$/\rho_0)=10.14$.
The sink radius is set to 2.5 grid cell lengths (of the maximally refined cells) to avoid artificial fragmentation.
Sink particles continue to accrete gas from any cells within the accretion radius of the sink particle that have exceeded $\rho$\textsubscript{sink} \citep{Federrath2015}.

The \texttt{All + Outflow} simulation uses a custom module for implementing a two-component jet feedback as described in \cite{Federrath+2014}. 
In the simulation with stellar feedback, each sink particle produces both fast collimated jets and wide-angle, lower-speed outflows. 
For more details about the sink particles in these simulations and the protostellar outflow prescription, see \cite{Federrath+2014, Federrath2015} and \cite{Appel2022}.  

Two snapshots, the 1\% \texttt{B-Fields} case and the 3\% \texttt{All + Outflow} case, are removed from all plots and analysis. 
The the 1\% \texttt{B-Fields} snapshot is removed because the PDF exhibits a significantly steeper power-law tail due to a stochastic star formation event resulting in a single massive sink particle \citep{Appel2022}. 
The 3\% \texttt{All + Outflow} case is removed because the PDF exhibits a strong second power-law tail.
We discuss these snapshots further in Appendix~\ref{app:outliers}.

We note that throughout the paper we use the integrated star formation efficiency (SFE) as a proxy for time (i.e., how evolved the run is). 
The integrated SFE measures how much of the initial gas mass has been converted into stellar mass (in the form of sink particles). 
Therefore, the integrated SFE can be used to characterize the evolutionary stage of each run, even when different physical modules are used, which can result in significantly different star formation timescales.

\section{Fitting Methods for Volume and Column Density PDFs}
\label{sec:models}

From the density field of each simulation snapshot, we constructed the volume-weighted PDFs of the volume density as well as the volume-weighted PDFs of the column density.
We generate the column density values by first making a covering grid of uniform resolution. 
We then project the density distribution along three separate lines of sight (LOS) which we have labeled the x, y, and z-axes. 
This results in a single volume density PDF and three different column density PDFs, corresponding to each LOS, for each snapshot. 
The fitted values are shown below for all three LOS, however, we only show the PDFs for a single, representative LOS.

In this section we describe our two methods for fitting the volume density PDFs ($\rho$-PDFs) and column density PDFs ($\Sigma$-PDFs) of our simulations.
We first discuss fitting the PDFs with a piecewise lognormal plus power-law distribution. 
We demonstrate the method presented in \cite{Khullar2021}, which uses non-linear least squares to fit a piecewise function to the PDFs. 
We then present our novel method of using a 5th-order spline polynomial to fit the full distribution of the PDFs.
We show that the derivatives can be used to find where the functional form of the PDF switches from a lognormal to a power-law, and to find the slope of the PDF as a function of density. 

\subsection{Piecewise Fitting method}  \label{sec:piecewise_method}

Our first approach, which we refer to as the piecewise fitting method throughout the paper, fits a piecewise distribution with a fixed analytical description to the density PDF.
We use the fitting method outlined in \cite{Khullar2021} and the corresponding fitting module available on \href{https://github.com/shivankhullar/PDF_Fit}{github}. 
This method uses a non-linear least-squares method to fit a pure lognormal (LN) distribution, a lognormal plus power-law (LN+PL), or a double power-law (LN+PLPL) distribution. 
Although a double power-law fit is possible with this fitting technique, we must choose the analytical form of the distribution (i.e., between a LN+PL and a LN+PLPL) before fitting the PDF. 
While most PDFs appear to have a LN+PL structure, there is evidence that some star-forming regions may have a double power-law structure (discussed further in Section \ref{sec:discussion}).

We consider the logarithm of the normalized density, 
\begin{equation}
    s = \ln(\rho/\rho_0 ) \, ,
\end{equation}
where $\rho_0$ is the mean density in the computational volume.
Then, the piecewise form of the LN+PL PDF is given by \citep{MyersP2015, Burkhart2017, Burkhart2018, BurkhartMocz2019}:
\begin{equation}
\label{eq:SPL}
p(s) =
\begin{cases}
\frac{N}{\sqrt{2 \pi \sigma_{\rm s}^2}} \exp\left[-\frac{(s-s_0)^2}{2 \sigma_{\rm s}^2}\right] & s< s_t, \\
N p_0 e^{-s \alpha} & s\geq s_t,
\end{cases}
\end{equation}
where $\alpha$ is the slope of the power-law tail, $\sigma_{\rm s}$ is the width of the lognormal, $s_0$ is the mean of the lognormal, and $N$ and $p_0$ are normalisation factors. 

The functional form described by Eq.~\ref{eq:SPL} depends on 6~variables, $N$, $p_0$, $s_0$, $\sigma_{\rm s}$, $s_t$, and $\alpha$.
The fitting method outlined in \cite{Khullar2021} imposes additional constraints on the fit, which result in only two independent variables ($\sigma_s$ and $\alpha$): 
\begin{enumerate}
    \item The integral over all values of $s$ of the PDF must be unity, i.e., $\int_{-\infty}^{\infty} p(s) ds = 1$.
    \item The PDF must be continuous everywhere, including at the transition density $s_{\rm t}$.
    \item The  PDF must be differentiable everywhere, including at the transition density $s_{\rm t}$ (i.e., the derivative, $dp/ds$ is continuous). 
    \item The total gas mass must also be conserved, i.e., $\int_{-\infty}^{\infty} e^{s} p(s) ds = 1$.
\end{enumerate}

This fitting method requires three user defined parameters: a maximum density cut off above which data is not included in the fit  (i.e., s$_{cut}$), an initial best guess for the width of the lognormal ($\sigma_s$), and an initial best guess for the slope of the power-law exponent ($\alpha$). 
The LN+PL fit function from the publicly available code described in \citep{Khullar2021} and utilized here outputs a best-fit value for both $\sigma_s$ and $\alpha$, along with errors based on the SciPy \verb|curve_fit| function \citep{2020SciPy-NMeth}. 
The errors are calculated by taking the square root of the diagonal of the covariance matrix produced by scipy's \verb|curve_fit| function. 

\begin{figure}
\centering
\includegraphics[width=
\linewidth]{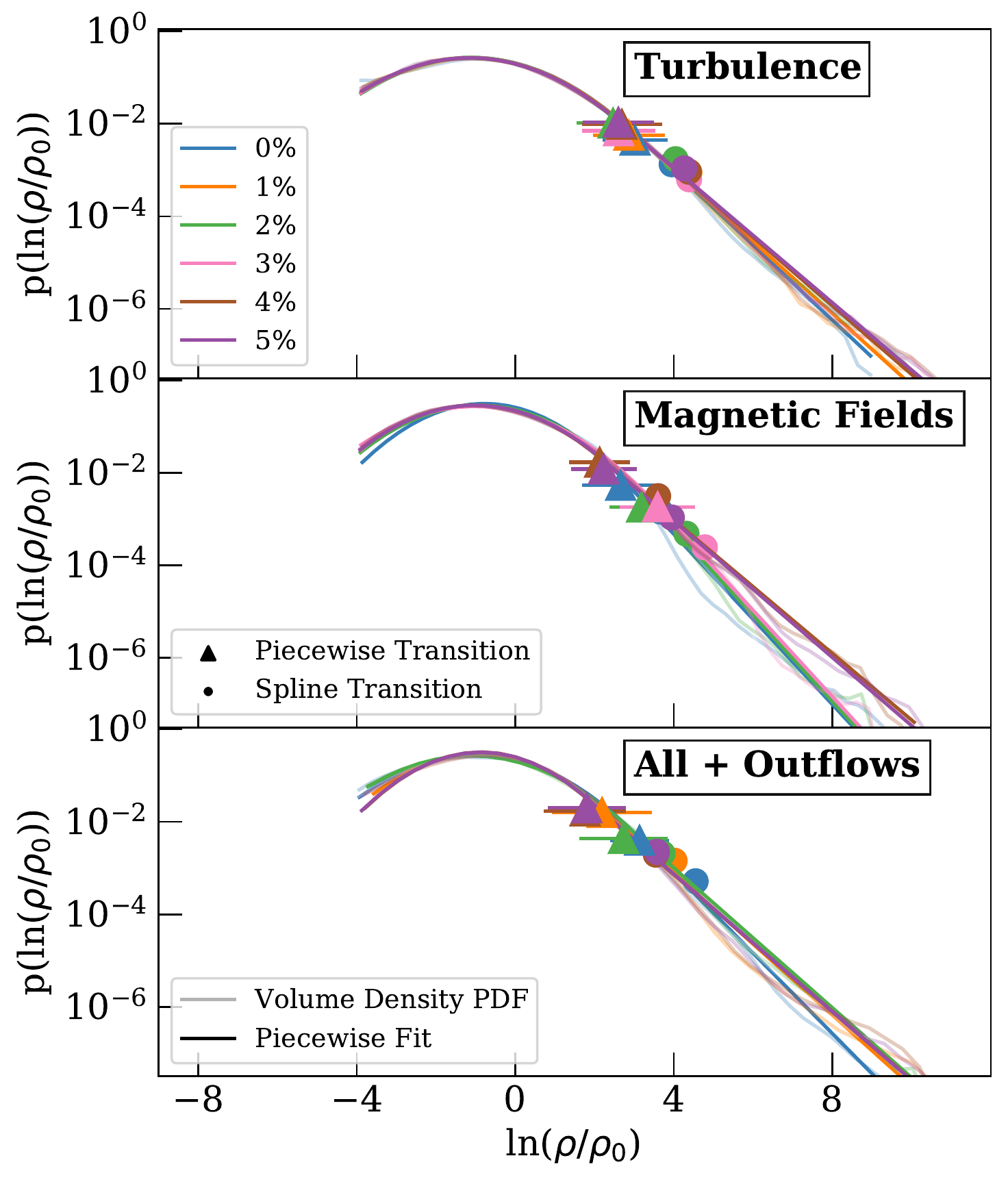}
\caption{Volume density PDFs for the \texttt{Turbulence} run (top), \texttt{B-Fields} run (middle), and \texttt{All + Outflow} run (bottom) shown with faint solid lines. We overplot the piecewise fits described in Section~\ref{fig:piecewise_fit} as bold solid lines. Snapshots from SFE=$0\%$ through $5\%$ are shown in different colors (the corresponding total SFE of each color is shown on the plot).  Colored triangle points show the location of the transition density ($s_t$) as estimated from the piecewise method, while colored circles indicate the transition density estimated from the spline fit. 
\label{fig:vol_piecewise}}
\end{figure}

The fitting method described above was used to fit the PDFs in Figs.~\ref{fig:vol_piecewise}~and~\ref{fig:piecewise_fit}. We display the fitted transition density from the piecewise fitting method as trianglular points.
More details on the assumptions of the piecewise fitting method used here can be found in \citet{Khullar2021} Appendix A.

\subsection{Spline Polynomial Fitting Method}  
\label{sec:spline}

\begin{figure*}
\centering
\includegraphics[width=.9
\linewidth]{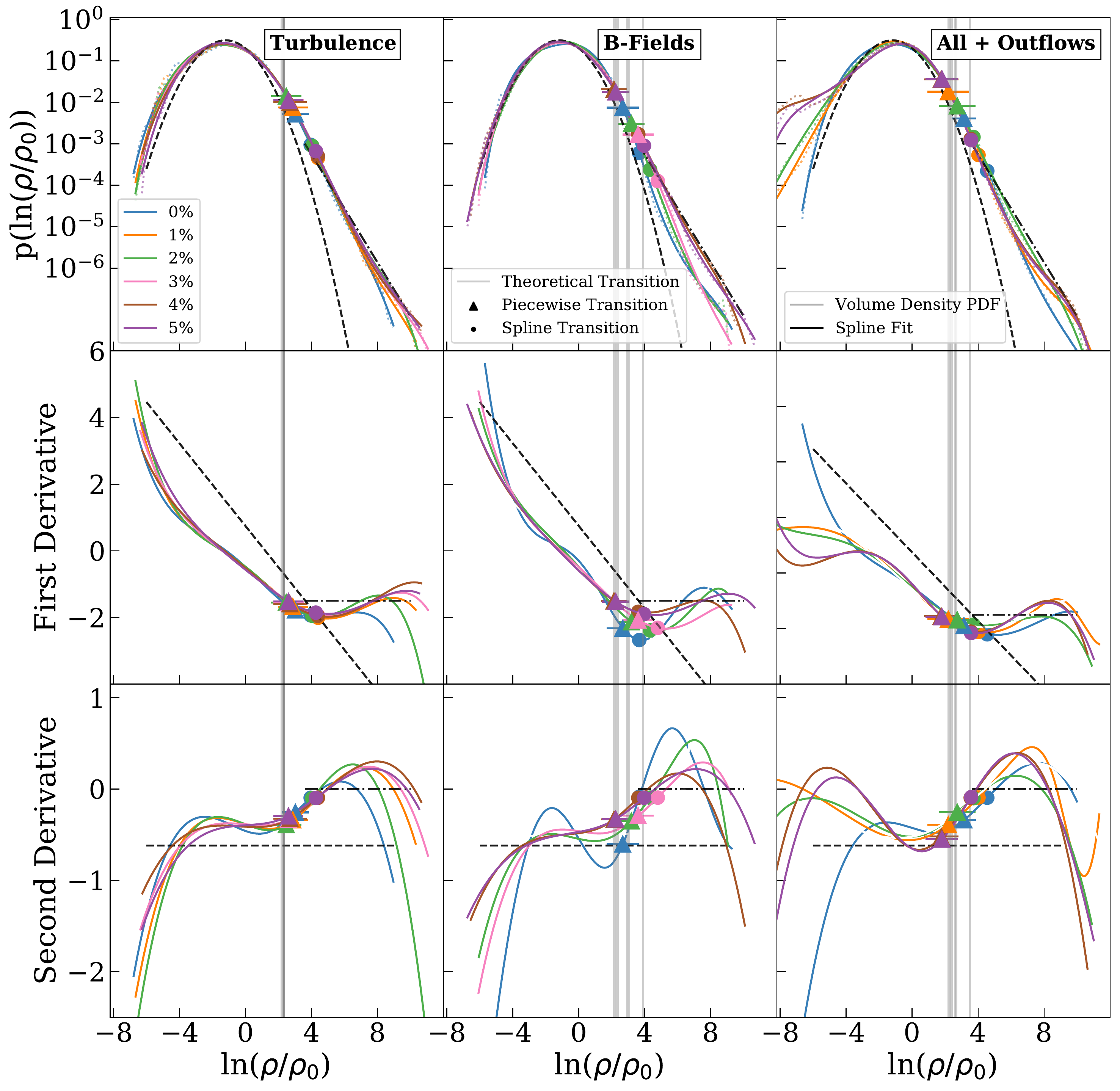}
\caption{Volume density PDFs for the SFE $=0\%$ through $5\%$ snapshots of each of the simulations described in Section~\ref{sec:sim} are shown in the top row. 
We overplot the spline fits described in Section~\ref{sec:spline}.
We also overplot a reference lognormal distribution with a width of  $s=1.27$ and a power-law distribution with a slope of $\alpha=-1.5$. 
The first and second derivative of each spline fit is plotted in the 2nd and 3rd row respectively, along with the derivative of the reference lognormal and power-law functions (black lines). The spline method helps to visualize the presence of two power-law tails by looking at where the second derivative becomes zero.
\label{fig:vol_spline}}
\end{figure*}

Our second method, which we refer to as the spline method throughout the paper, fits the PDF with a 5th-order spline polynomial. The default for spline is a 3rd-order polynomial. We find no substantial difference in our results with a higher order polynomial, and choose a 5th-order to make our figures clearer.
We fit each PDF with SciPy's \verb|UnivariateSpline| function \citep{2020SciPy-NMeth} and analyze the spline fit's first and second derivatives to investigate how the functional form of the PDF changes with density.
In particular, we use the \verb|UnivariateSpline| function from the \verb|scipy.interpolate| package, and interpolate a fit to the PDFs using a 5th degree polynomial. 
\verb|UnivariateSpline| uses a $k$ value (degree of polynomial) of $k=3$ as a default. We use  $k=5$ to get our 5th degree polynomial. We set the smoothing factor (S), to $S=.3$. The smoothing factor has no substantial effect on our results.
\verb|UnivariateSpline| is a function that interpolates a piecewise-polynomial at a given order and given smoothing factor.
The derivatives of the spline are useful for characterizing the PDF and can be used to measure the transition density from lognormal to power-law forms, determine if there is more than one power-law, and find the slope as a function of density.
For instance, the spline method computes the transition density as the density value at which the second derivative becomes zero (as is expected for a purely linear distribution). 
We take the value of the spline fit's first derivative at this point as the power-law slope. 
The spline fit, as well as the first and second derivatives, can be seen in Figs.~\ref{fig:vol_spline}~and~\ref{fig:spline_fit}. 
We set a threshold near zero of $0.1$ to determine the transition to power-law tail, because the stochastic fluctuations of the underlying PDF means the second derivative may not be exactly zero. The filled circles represent the transition density as measured from the spline fit.

\section{Results} 
\label{sec:results}

\subsection{Volume Density PDFs} \label{sec:volume}

We show the $\rho-$PDFs in Fig.~\ref{fig:vol_piecewise} for all three of our simulations. 
The top panel contains the volume density PDFs for the \texttt{Turbulence} simulation described in Section~\ref{sec:sim}, displayed as faint lines. 
The middle panel shows the same for the \texttt{B-Fields} simulation, and the bottom panel shows the \texttt{All + Outflow} simulation. 
We do not fit densities below $s=-4$ for all three cases, as these lowest densities are subject to strong fluctuations due to feedback effects in the \texttt{All + Outflow} simulation \citep{Appel2022}.
Each color represents a different SFE from  SFE $=0\%$ through $5\%$. 
The opaque lines represent the piecewise fit of the PDFs using the \citet{Khullar2021} fitting algorithm described in Section~\ref{sec:piecewise_method}. 

Visually, the single LN+PL piecewise fitting method fits the simulated volume density PDFs well. 
The exception to this is in the case of the \texttt{All + Outflow} simulation, which are not fit well by a PL fit at densities past $s=6$.
The \texttt{Turbulence} simulation shows the most time stable power-law slope, while the \texttt{B-Fields} simulation has power-law slopes that vary significantly over time (i.e., vary between snapshots of different SFE).  

The fitted transition densities from the piecewise fitting method (trianglular points) are roughly the same for all three simulation runs. 
\citet{BurkhartMocz2019} showed that the transition density between the lognormal and the power-law forms depends on the strength of the turbulence in the cloud, as described by the sonic Mach number and the virial parameter. 
Given that each of our simulations have turbulence driven in the same way, it is not surprising that the simulations have similar transition densities.

The spline method transition densities (the circles in Fig.~\ref{fig:vol_piecewise}) all lie at slightly higher densities than the transition densities measured using the piecewise fitting method.  
We show the fits and the first and second derivatives from the spline fitting method for the volume density PDFs in Fig.~\ref{fig:vol_spline}. 
The top row is similar to Fig.~\ref{fig:vol_piecewise} and shows the density PDFs (faint, dotted lines) and transition densities of both fitting methods.
Figure~\ref{fig:vol_spline} also shows the fitted polynomial spline functions (solid lines).
We include a reference lognormal distribution (black dashed curve) with a width of $\sigma=1.27$ and a reference power-law distribution (black dot-dashed line) with a slope of $\alpha=-1.5$.
The middle row shows the first derivative of both the spline fits and the two reference distributions. 
The bottom row shows the second derivatives of the fits and the two functions.
All panels mark the transition density from the spline method (circles) and the piecewise fit method (triangles). 
Figure~\ref{fig:vol_spline} also shows the theoretical transition densities (grey vertical lines), calculated using the power-law slope ($\alpha$) measured using the spline fitting method, the lognormal width ($\sigma_s$) from \cite{Appel2022}, and Equation~8 from \cite{Appel2022}. 

The advantage of fitting a polynomial spline and using its derivatives to determine when the PDF function changes from a lognormal distribution to a power-law distribution is apparent upon examination of the reference distributions. Since the Spline is a piecewise polynomial, we can calculate its derivatives analytically at each point. 
These derivatives can then be compared to the derivatives of the reference functions.
The pure lognormal function has a linear first derivative and a constant valued (non-zero) second derivative. 
The power-law function has a first derivative that is constant and equal to the slope of the power-law, and a second derivative that is zero. 
We therefore determine the transition to a power-law distribution as the point at which the spline fit flattens out in the first derivative and the second derivative approaches zero.
We use the 2nd derivative to measure this point and indicate it on the plot with a circular point for each snapshot. 

While the simulated PDFs display significantly more complexity than the idealized lognormal and single power-law functions in the shapes of their derivatives, their overall behavior follows the expected transition between the two functions. 
In the lognormal portion of the density PDF, the first derivatives linearly decline as the slopes of the PDF shift from positive to negative at roughly the mean density. 
This linear decrease levels off and the first derivative becomes roughly constant at the transition density to a power-law form (around $s=3$, with some variation between snapshots and simulations).  

An additional advantage of fitting the spline is that it does not rely on user choice in determining if the form of the fitted PDF is a single or double power-law. 
In many cases, the PDF spline derivatives clearly show the presence of a second power-law toward higher density (around $s=5-6$). This can especially be seen in the \texttt{All + Outflow} case, in which several curves could be fit by a double PL past $s=6$.  
The slopes of the power-laws can be determined from the first derivative plots. 
The first power-laws have a slope around $-2$ and the second power-laws are significantly shallower (around $-1.5$). 
This second power-law would be entirely missed by fitting a piecewise function with a single power-law, and highlights the utility of a model-independent fitting method, such as a spline, which is agnostic to model constraints, and a user-defined fitting function, if the polynomial order chosen is sufficiently high.

\begin{figure}
\includegraphics[width = 
\linewidth]{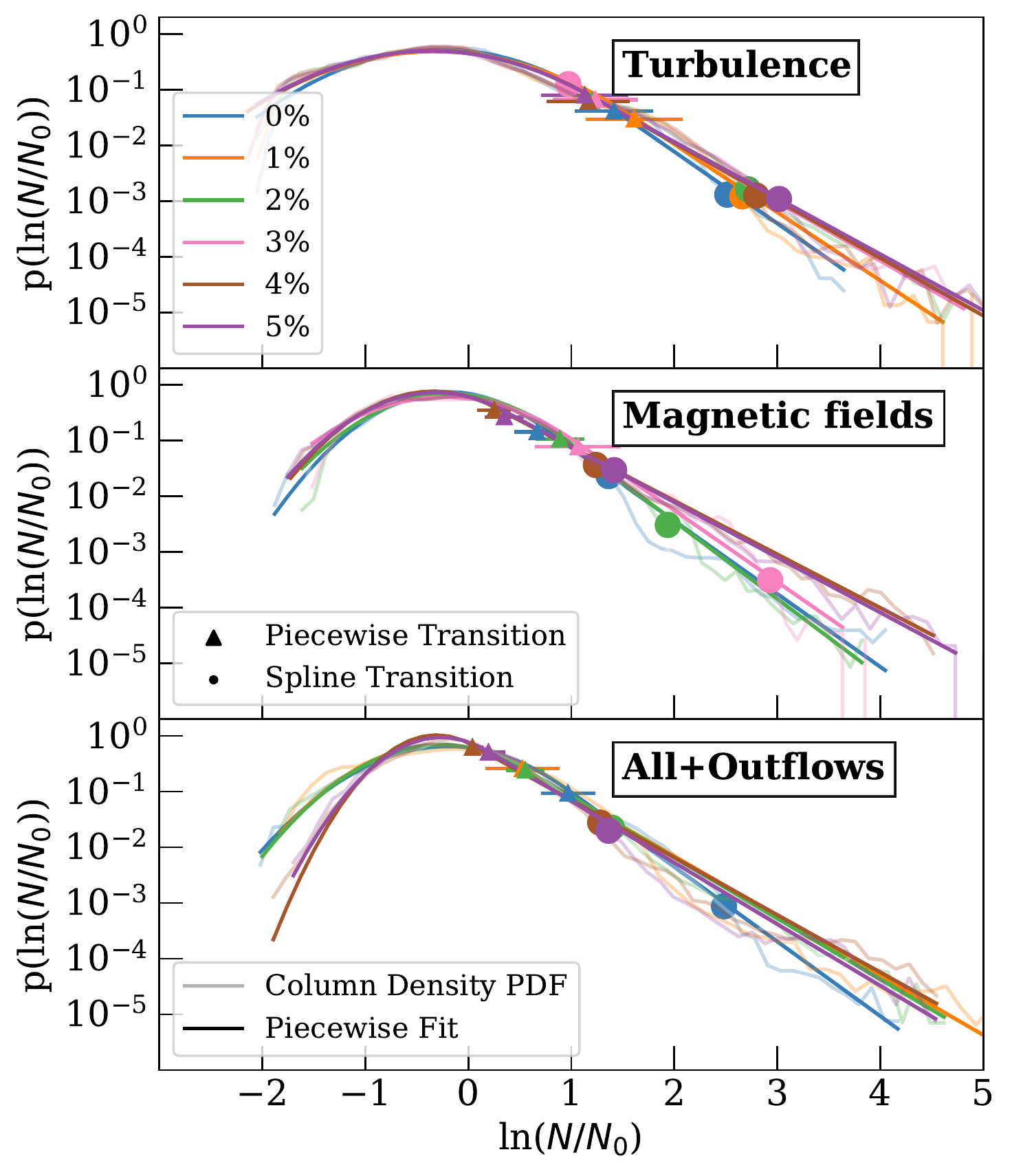}
\caption{The same as Fig.~\ref{fig:vol_piecewise} but for the column density PDFs rather than volume density PDFs.
}
\label{fig:piecewise_fit}
\end{figure}


\begin{figure*}
\centering
\includegraphics[width = 0.9\linewidth]{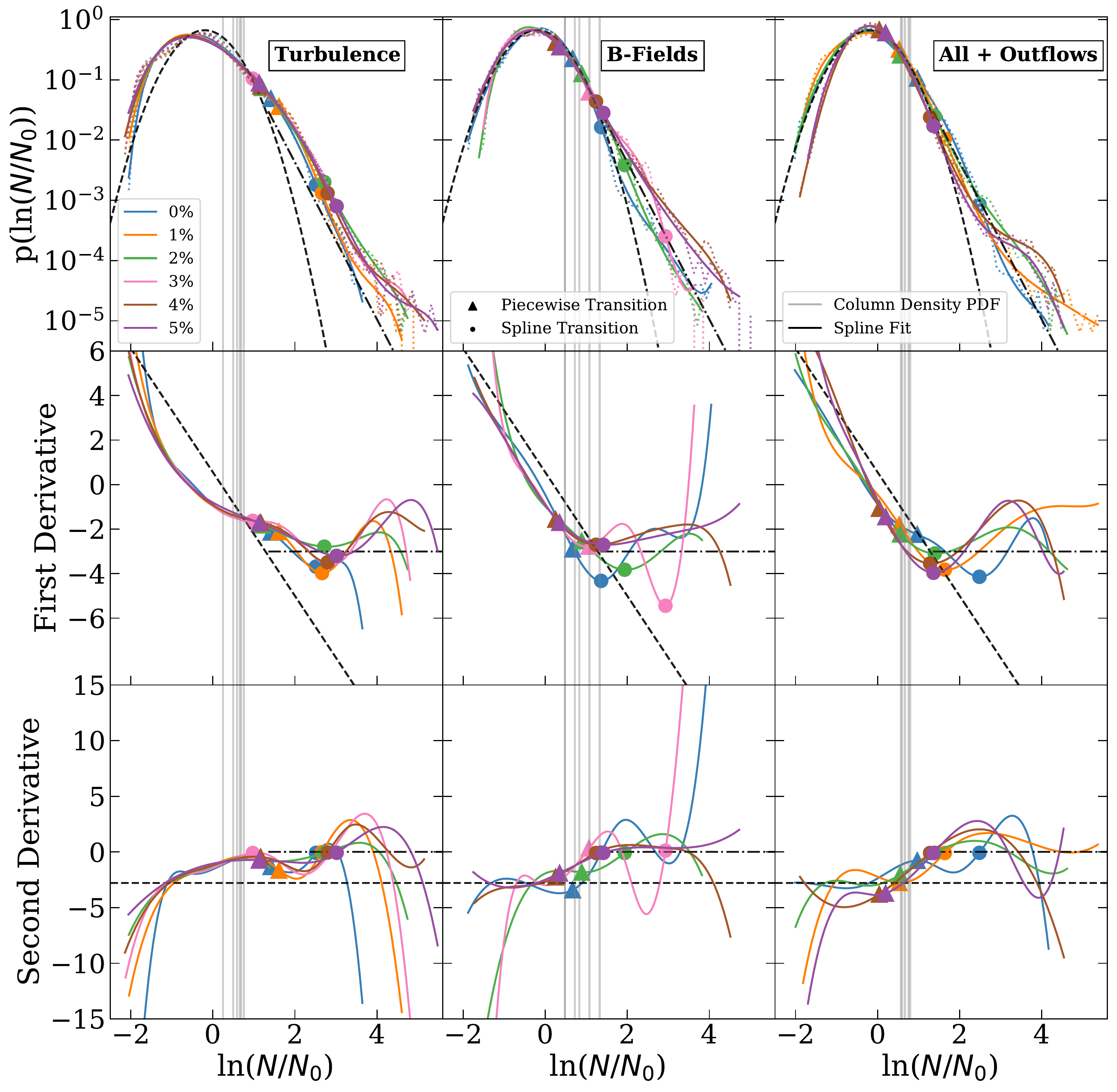}
\caption{The same as Fig.~\ref{fig:vol_spline} but for the column density PDFs rather than the volume density PDFs. We also overplotted a reference lognormal distribution with a width of $\sigma_s=0.6$ and power-law with a slope of $\alpha=-3$ based upon averages from analysis. 
\label{fig:spline_fit}}
\end{figure*}


\subsection{Column Density PDFs}

Despite being important for theories of star formation, volume density PDFs are not directly observable. 
The column density, on the other hand, can be observed, making the column density PDF much more readily available from observational data.
However, there are additional challenges to using column density PDFs. 
First, in order to compare to analytic models of star formation, we are interested in measuring the transition densities and power-law slopes of the volume density PDFs -- which means we must convert from the transition density and slope of the observed column density PDF to the corresponding volume density transition density and slope. 
We discuss this conversion in Section~\ref{sec:column_volume}.
In addition, column density PDFs are significantly more noisy than volume density PDFs and, hence, it may difficult to distinguish if a single power-law tail or a double power-law tail is the most appropriate choice of functional form for a fit. 
Therefore, having a fitting method that does not depend on choosing the function of the fitted curve, such as fitting a spline and using the derivatives to determine slopes and transition densities, is useful. 

We apply both of the fitting methods discussed above to the column density PDFs of our simulations, as described in Section~\ref{sec:models}.
As with the volume density PDFs, we first investigate a LN + PL fit using the \citet{Khullar2019} piecewise fitting method, as shown in in Fig.~\ref{fig:piecewise_fit}.
We show a single line of sight along the direction of the mean magnetic field as we find our results do not depend on the chosen LOS. 
The faint lines show the column density PDFs and the solid lines show the fits of a lognormal plus single power-law model.  
The column density PDFs are statistically much noisier than volume density PDFs for our simulations. 
In observational data, additional noise, beam smoothing effects, radiative transfer effects, and selection effects may also bias and distort the PDF shapes \citep{Ossenkopf02a, Lombardi2015AA, schneider15, Alves2017AA}.

The measured transition densities of the column density, as determined by both the fitted piecewise function (trianglular points) and the spline method (circlular points), are more spread out than the volume density transition densities. 
This may be due to the fact that column density PDFs exhibit far more fluctuations in higher column density bins than the high-density end of the volume density PDFs, due to low-number statistics. 
We also note that the very low column density LN widths vary significantly for the \texttt{All + Outflow} simulation as compared to the other two runs, which have roughly constant LN widths. 
This is likely due to the presence of protostellar outflows, which drive additional turbulence in the numerical box \citep{Appel2022,Hu2022,Appel2023}.

Figure \ref{fig:spline_fit} shows the spline fitting method for the column density PDFs. 
As described in Section ~\ref{sec:volume}, Fig.~\ref{fig:spline_fit} contains reference lognormal and power-law distributions, but here we use a width of $\sigma=0.6$ and a slope of $\alpha=-3$, respectively (in accordance with the expectations for the column density PDF from \citealt{Burkhart12}). 
As was the case for the volume density PDFs, the spline method indicates the presence of multiple power-law tails in the column density PDFs, which would be missed if using a piecewise fit with only a single power-law tail. This can be see in Figure \ref{fig:spline_fit} as the 0\% and 3\% cases in \texttt{B-Fields} has a prominent 2nd power-law tail.

\subsection{Comparing Slopes Estimated from the Piecewise Fit and the Spline Fit}

\begin{figure}
\centering
\includegraphics[width = \linewidth]{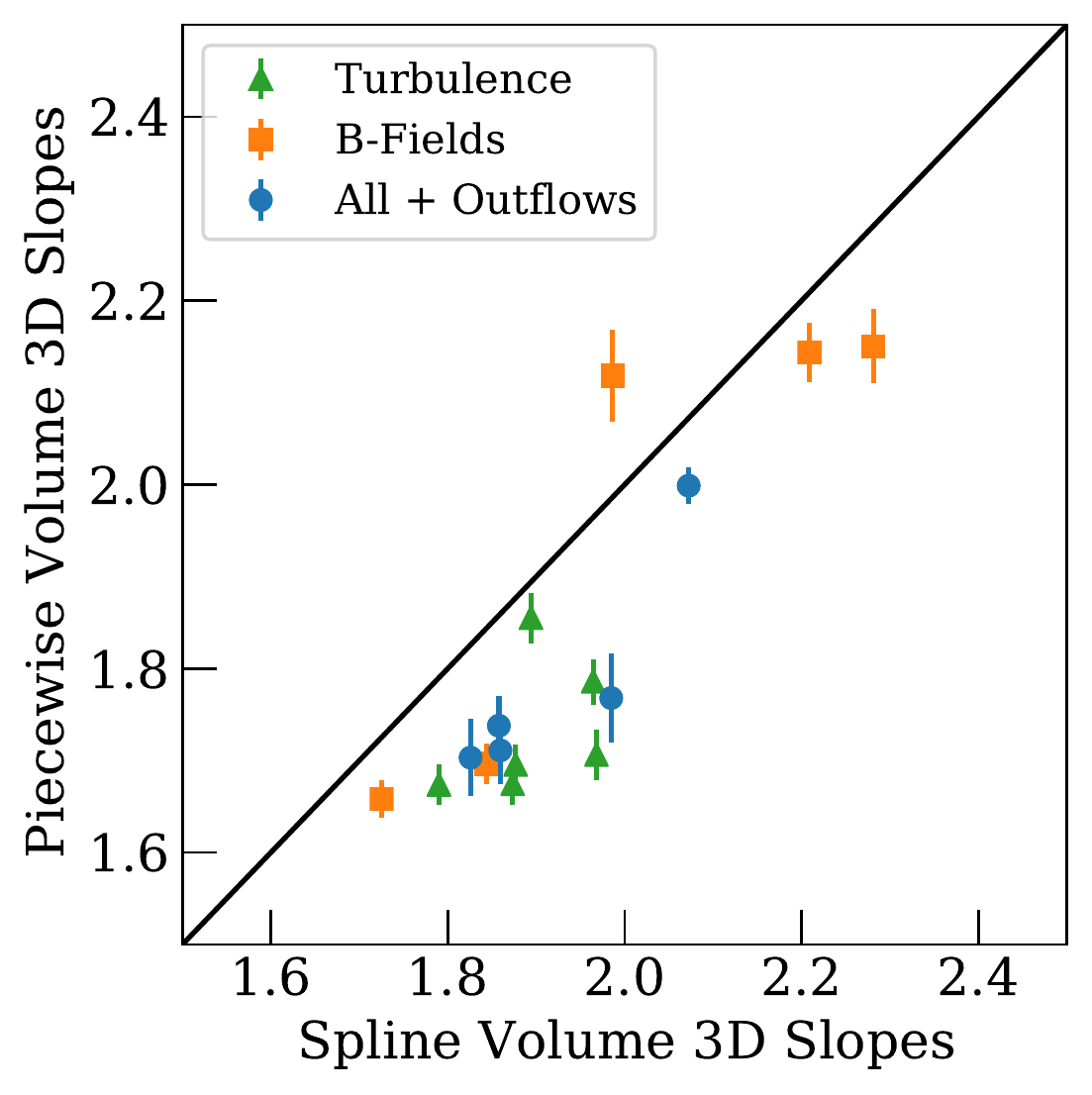}
\caption{A comparison of the fitted slopes of the volume density PDFs ($\alpha_{s}$) calculated from the piecewise fitting method versus the spline fitting method. The black diagonal line shows the one-to-one line. Error bars are shown for the piecewise fitting method as described in Section \ref{sec:piecewise_method}.
\label{fig:fits_3D_comparision}}
\end{figure}

\begin{figure}
\centering
\includegraphics[width = \linewidth]{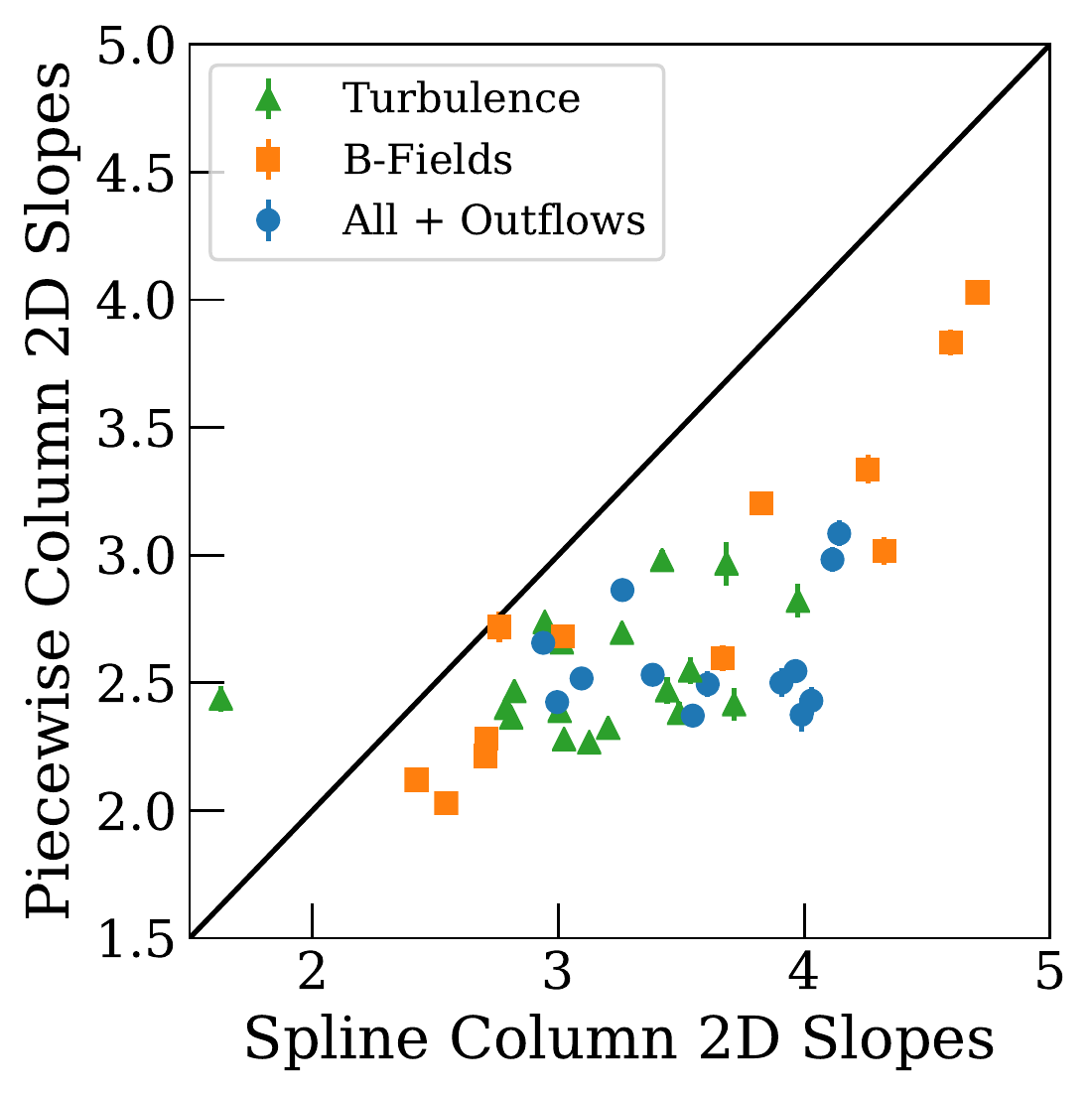}
\caption{Same as Fig.~\ref{fig:fits_3D_comparision} but for the column density PDFs ($\alpha_{\eta}$) from the piecewise fitting method versus the spline fitting method. Error bars are shown for the Piecewise fitting method as described in Section \ref{sec:piecewise_method}.
\label{fig:fits_2D_comparision}}
\end{figure}

The slopes produced by the two different fitting methods can also be used to directly compare the methods, as can be seen in Fig.~\ref{fig:fits_3D_comparision} for the volume density PDFs and Fig.~\ref{fig:fits_2D_comparision} for the column density PDFs. 
We only consider the first power-law slope value found from the spline method.
Figures~\ref{fig:fits_3D_comparision}~and~\ref{fig:fits_2D_comparision} directly compare the fitted slope values produced for a specific volume or column density PDF from the piecewise and the spline fitting methods. 
Figure~\ref{fig:fits_3D_comparision} shows that both methods are in reasonable agreement for volume density slope values. 
However, as can be seen in Fig.~\ref{fig:fits_2D_comparision}, the spline method produces steeper slope values for the column density PDF than the piecewise method does. 
This is a result of the fitted transition density falling at a higher density value in the spline method than the piecewise method.

The piecewise method consistently fits the transition density at lower densities than the spline method, and at densities where the first derivatives are still negative and have not yet gone to zero.
As implied by our investigation, the transition density can be thought of as more like a range, in which the PDF leaves a lognormal distribution and transitions into a power-law distribution. The piecewise method is finding the transition density as the moment the lognormal starts turning into a power-law, whereas the Spline is finding the transition density as the point at which the PDF becomes a true power-law. This is because the spline method is looking at the derivative of the curve to find the transition density the moment the PDF becomes a true power-law.

\subsection{Connecting the Column Density PDFs to the Underlying Volume Densities PDFs} 
\label{sec:column_volume}

\begin{figure*}
\centering
\includegraphics[width = .8 \linewidth]{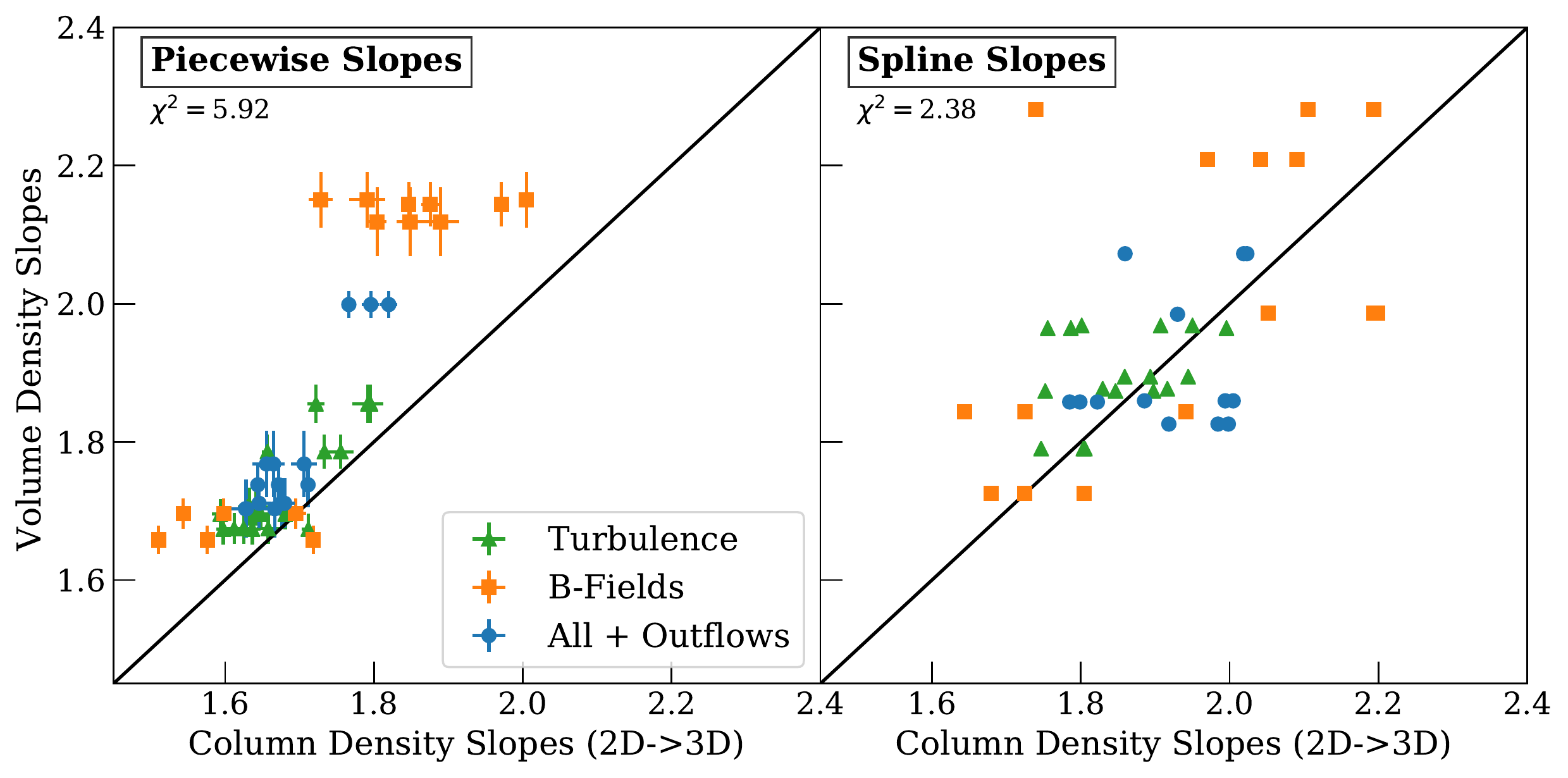}
\caption{\textbf{Left:} The fitted slopes of the volume density PDFs from the piecewise fitting method is shown  on the y-axis. The x-axis represents the corresponding fitted slopes of the column density PDFs converted to a 3D slope value using Eq.~\ref{eq:slope_conversion}. \textbf{Right:} The same plot, but for the spline fitting method. Each panel shows the slopes for all three simulations, with 6 snapshots each, for the 0$\%$ through 5$\%$ SFE snapshots, excluding the outliers mentioned in Section \ref{sec:sim}.
\label{fig:slope_3D_comparision}}
\end{figure*}


Testing analytic star formation theories that use the volume density PDF requires converting properties of the column density PDF into the corresponding properties of a volume density PDF. 
To this end, \cite{Federrath+Klessen2013} proposed a formula to convert column density PDF slopes to volume density PDF slopes \citep{Federrath+Klessen2013, JaupartChabrier2020}:
\begin{equation}
    \alpha_\eta=\frac{3}{\frac{2}{\alpha_s}+1},
    \label{eq:slope_conversion}
\end{equation}
where $\alpha_\eta$ is the slope of the power-law tail for volume density, and $\alpha_s$ is the slope of the power-law tail for column density.
Using our fitted slopes from both fitting methods for the first power-law, we compare the converted column density slopes to the volume density slopes.
In other words, we use Eq.~\ref{eq:slope_conversion} to convert the fitted slopes of the column density PDFs to volume density slopes.
The comparison between the fitted volume density slopes and the converted column density slopes is shown in Fig.~\ref{fig:slope_3D_comparision}. 
We perform this comparison for both the spline and piecewise fitting methods. 
The solid black line in each panel of Fig.~\ref{fig:slope_3D_comparision} shows the one-to-one line.

Most of the points in the left panel of Fig.~\ref{fig:slope_3D_comparision} lie above the one-to-one line. 
This suggests that the piecewise method tends to produce column density slopes that, when combined with Eq.~\ref{eq:slope_conversion} (i.e., the x-axis), under-predict the true volume density slope values (i.e., the y-axis). 
In contrast, the spline method produces predicted volume density slopes that have a greater spread but are more centered on the one to one line (see the right panel of Fig.~\ref{fig:slope_3D_comparision}). 
We calculate a $\chi^2$ value for both panels of Fig.~\ref{fig:slope_3D_comparision} in order to quantify which fitting method resulted in a more reliable conversion of column density slopes to volume density slopes using Eq.~\ref{eq:SPL}. 
The piecewise fitting method has a chi-squared value of $\chi^2=5.92$ and the spline fitting method has a chi-squared value of $\chi^2 = 2.38$.  
This indicates that the spline fitting method produces a more consistent PL slope between volume and column density versions of the PDF.


\section{Discussion}
\label{sec:discussion}

Molecular cloud density probability distribution functions (PDFs) are important tools for understanding the structure and evolution of molecular clouds. In particular, the column density PDF, which describes the distribution of column densities within a molecular cloud, can provide valuable insights into the physical processes at work within the cloud. For example, the transition from a lognormal to a power-law form is thought to be a key indicator of the gravitational instability of the cloud and can measure the self-gravitating gas fraction \citep{BurkhartMocz2019,Kritsuk+2011,Federrath+Klessen2013,Girichidis2014}. The dynamics of the gas residing in the power-law is dominated by self-gravity \citep{Appel2022}. The fraction of gas in the power-law has been shown to be a good tracer of the SFE in both simulations and observations \citep{Burkhart2018,bemis2023}.

Fitting a model to the volume density or column density PDF allows us to determine the underlying physical processes that give rise to the observed distribution. This is typically done using a best-fit approach, which compares the observed data to a model and quantifies the goodness of fit. However, this approach has limitations, as it is sensitive to the choice of binning \citep{Brout2021}, can be affected by noise in the data, and ultimately requires human-choice in the fitting model function \citep{Khullar2021}.

Our work provides a new way of determining the transition point between LN and PL forms and hence the onset of the dominance of self-gravity in molecular clouds. By fitting a spline function to the PDF, we are able to examine the PDF's 1st and 2nd derivatives and therefore determine where the function  transitions from LN to PL. Furthermore, we can determine if there are multiple PLs at the highest densities without having to make a human-based decision on fitting single or multiple PL functions. We also can directly read off the slopes of the PL from the value of the first derivatives. 

 The spline method is particularly useful as we have shown that piecewise best-fit methods can under-predict the transition density in our column density PDFs. This under-prediction leads to shallower slopes in the column density PDF, which then gives rise to an offset in the conversion between column density slopes and volume density slopes (i.e., Fig.~\ref{fig:slope_3D_comparision}, left panel).  If the PDF is less noisy, this is less of an issue. For example, in our density PDFs the predicted transition densities and power-law slopes from both fitting methods are similar.

Close inspection of the triangular points (transition densities fitted by the piecewise fitting method) in spline plots of Figs.~\ref{fig:vol_spline}~and~\ref{fig:spline_fit} reveal that the piecewise fitting method tends to place the transition point where the first derivative curve still functionally looks like the idealized LN function (i.e., a straight line). In other words, the piecewise method tends to fit the transition at a lower density than the function's derivative would suggest. This could be due to the additional constraints that the \cite{Khullar2021} fitting method utilizes (see Section~\ref{sec:models}) and can explain the offsets in Fig.~\ref{fig:fits_2D_comparision}.

\subsection{Comparison to Observations}

Although analytic models of star formation often rely on the volume density PDF, actually observing the volume density distribution is incredibly difficult (although some work is beginning to address this; e.g., \citealt{Dharmawardena2022}). 
However, many observations of the column density distribution of molecular clouds have been made and significant work has been done to characterize the shape of these column density PDFs for these clouds \cite[see e.g.,][]{Schneider+2015,Schneider+2016,Alves2017AA, Ma+2021, Ma+2022}.

\cite{Ma+2022} show that, although many observed molecular clouds have a purely lognormal $\Sigma$-PDF, 60\% of clouds with a LN+PL $\Sigma$-PDF show evidence of active star formation. 
\cite{Schneider+2016} also observe star-forming clouds with a LN+PL distribution.
Indeed, \cite{Schneider+2015} further show evidence of column density PDFs with two power-law tails for high-density star-forming clumps.

\cite{Schneider+2015} suggest a few possible interpretations of the second power-law, including rotation effects, thermodynamical effects, and magnetic fields. 
\cite{Khullar2021} also suggest that the second power-law may be due to rotational effects, i.e., accretion disk formation.
However, the ultimate cause of this second power-law will require further investigation.
Critically for our work, the variety in the shapes of the observed $\Sigma$-PDFs (i.e., LN, LN+PL, LN+PLPL) means that fitting methods that do not assume the functional form of the PDF are critical in order to find adequate fits to the observed PDFs.
Indeed, the shape of the PDF itself may serve as an important indicator of properties of the molecular cloud (e.g., whether the cloud is star-forming or not; \citealt{Ma+2022}).


\section{Conclusions}
\label{sec:conclusions}

Measuring the transition density between the lognormal and power-law portions of column and volume density PDFs, as well as the slope of the power-law tail, is important for relating the gas dynamics to the star formation rate in molecular clouds. 
However, methods for fitting the PDF can be prone to errors and to biases (e.g., the user's choice of a fitting function). 
In this paper, we compared two different fitting methods for the PDFs of numerically simulated star-forming molecular clouds.  
We used a series of three 3D hydrodynamic simulations with varying degrees of physics included. 
The two fitting methods that we compare, are using a best-fit of a piecewise LN+PL function to measure the power-law slope and the transition density between the lognormal and power-law forms of the PDF, and fitting a spline function and examining its derivatives to determine the slopes and transition densities. In summary, we find:
\begin{itemize}

\item Fitting a spline function to measure properties of the PDF removes reliance on the user's choice of fitting function and therefore allows us to directly determine if the data has multiple power-law slopes. 

\item The first power-law (set by the transition between lognormal and power-law function) can be visualized in the spline derivatives directly as the point where the first derivative flattens out and the second derivative goes to zero. 

\item The transition densities produced with the spline fitting method tend to be at higher values than those produced with the piecewise fitting method, which in turn also affects the measured slope. In general, the two methods produce fits that agree reasonably well for the volume density PDFS but that vary for the column density PDFs. 
This is likely due to the increased statistical noise in the column density maps as compared to the volume density. 

\item In testing a relation for estimating the volume density power-law slopes from the column density slopes, we find that the spline method produces a better match ($\chi^2$ of 2.38 vs.~$\chi^2$ of 5.92), albeit with significant scatter.

\end{itemize}

\acknowledgements

We thank Ana Gabela for her careful read of the manuscript. B.B.~thanks the the Alfred P. Sloan Foundation and the Packard Foundation for generous support. B.B., S.A., and A.K.~acknowledge the support of NSF grant AST-2009679. B.B.~is also supported by NASA grant No.~80NSSC20K0500.
The authors acknowledge the Office of Advanced Research Computing (OARC) at Rutgers, The State University of New Jersey for providing access to the Amarel cluster and associated research computing resources that have contributed to the results reported here
V.S. gratefully acknowledges the support provided by Harvard University through the Institute for Theory and Computation Fellowship.
C.F.~acknowledges funding provided by the Australian Research Council (Future Fellowship FT180100495 and Discovery Projects DP230102280), and the Australia-Germany Joint Research Cooperation Scheme (UA-DAAD). C.F.~further acknowledges high-performance computing resources provided by the Leibniz Rechenzentrum and the Gauss Centre for Supercomputing (grants~pr32lo, pn73fi, and GCS Large-scale project~22542), and the Australian National Computational Infrastructure (grant~ek9) in the framework of the National Computational Merit Allocation Scheme and the ANU Merit Allocation Scheme. Avery would like to thank their cat, Andromeda, for providing emotional support throughout the process of researching for and writing this paper.

\begin{figure*}
\centering
\includegraphics[width = 0.8\linewidth]{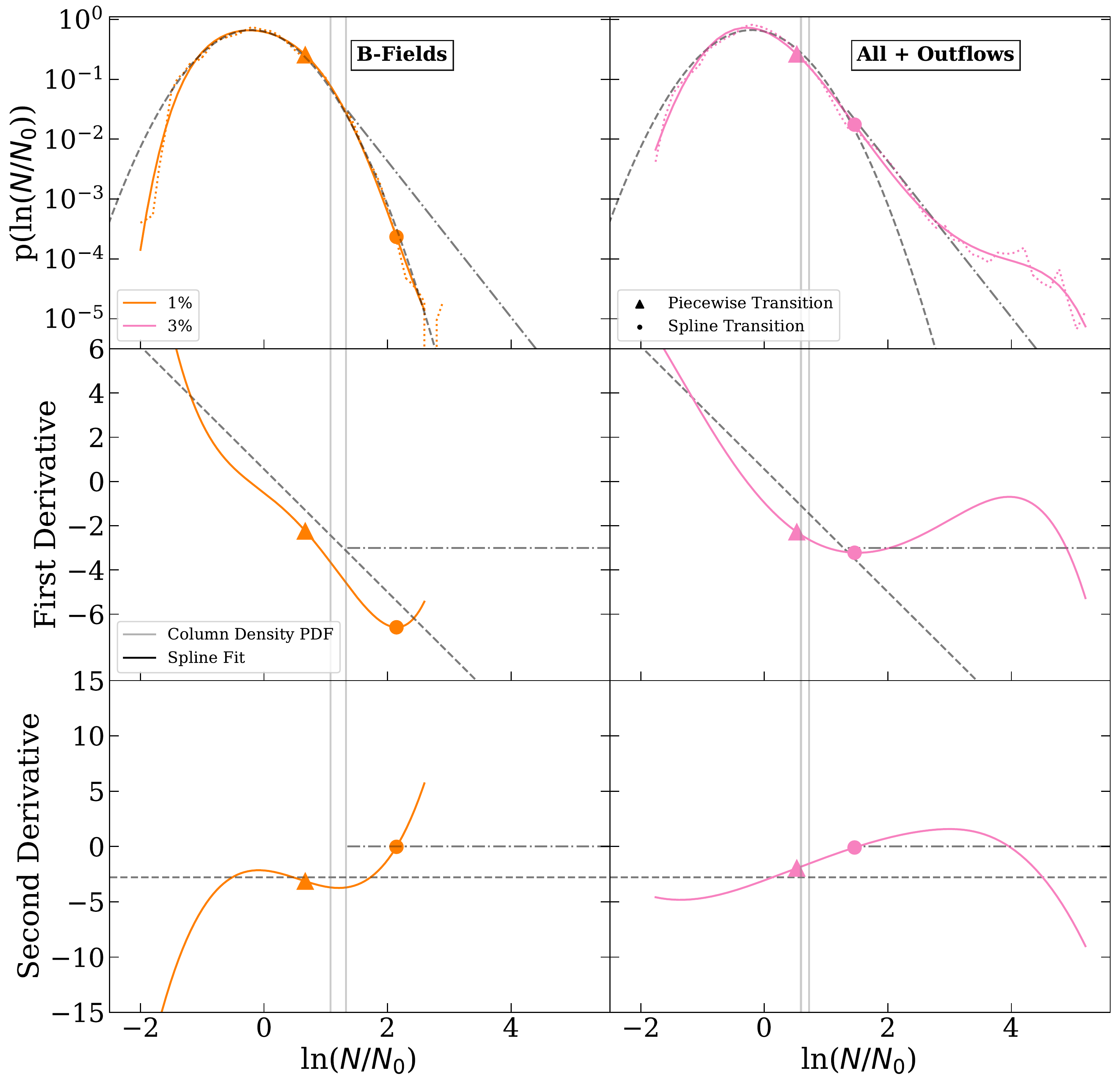}
\caption{The same as Fig.~\ref{fig:spline_fit} but for the column density PDFs of the two snapshots that where removed (the 1\% \texttt{B-Fields} case and the 3\% \texttt{All + Outflow} case). Also shown are a reference lognormal distribution with a width of $\sigma_s=0.6$ and power-law with a slope of $\alpha=-3$ based upon averages from analysis. 
\label{fig:outliers}}
\end{figure*}

\appendix 

\section{Removed Snapshots} \label{app:outliers}

In this section we further explore the two snapshots that were removed:the 1\% \texttt{B-Fields} case and the 3\% \texttt{All + Outflow} case.
Figure~\ref{fig:outliers} shows the column density PDFs using the spline fitting method for the 1\% \texttt{B-Fields} and the 3\% \texttt{All + Outflow} snapshots.

The 1\% \texttt{B-Fields} case in the top left panel of Fig.~\ref{fig:outliers} closely follows the example lognormal distribution and lacks a power-law tail. 
This is due to the fact that at just before this snapshot is saved, a single massive sink particle forms \citep{Appel2022}. We remove this snapshot because we are specifically exploring fitting PDFs with LN+PL distributions. 

The 3\% \texttt{All + Outflows} case is shown in the top right panel of Fig.~\ref{fig:outliers} and exhibits a strong second power-law tail. 
Although the spline method does not require selecting a particular distribution, our piecewise fitting method requires selecting either a single or a double power-law distribution and our current analysis focused on fitting a single power-law tail.
This case cannot be fit with any confidence with a single power-law tail and so we remove this snapshot from our analysis.

\bibliographystyle{aasjournal}

\bibliography{bibliography.bib,vs.bib}{}

\end{document}